\documentclass[preprint,12pt,authoryear]{elsarticle}

\usepackage{amssymb}

\usepackage{lineno}

\usepackage{booktabs}
\usepackage{acronym}
\usepackage{hyperref}
\hypersetup{
    colorlinks,
    linkcolor={black},
    citecolor={black},
    urlcolor={black}
}

\acrodef{AEA}{American Economic Association}
\acrodef{AJPS}{American Journal of Political Science}
\acrodef{API}{application programming interface}
\acrodef{CASD}{Centre d'accès sécurisé de données}
\acrodef{CJE}{Canadian Journal of Economics}
\acrodef{CSWEP}{AEA Committee on the Status of Women in the Economics Profession}
\acrodef{DCAP}{data and code availability policy}
\acrodef{DOI}{Digital Object Identifier}
\acrodef{DUA}{data use agreement}
\acrodef{EIA}{Energy Information Administration}
\acrodef{EJ}{Economic Journal}
\acrodef{FAIR}{Findable, Accessible, Interoperable, Re-usable}
\acrodef{FAQ}{frequently asked questions}
\acrodef{FSRDC}{Federal Statistical Research Data Centers}
\acrodef{GDPR}{General Data Protection Regulation}
\acrodef{GSOEP}{German Socio-Economic Panel}
\acrodef{HPC}{high performance computing}
\acrodef{HRS}{Health and Retirement Study}
\acrodef{IAB}{Institute for Employment Research in the German Federal Employment Agency}
\acrodef{IAB}{Research Data Center (FDZ) at the Institute for Employment Research}
\acrodef{ICPSR}{Inter-university Consortium for Political and Social Research}
\acrodef{IRB}{institutional review board}
\acrodef{JASA}{Journal of the American Statistical Association}
\acrodef{LEHD}{Longitudinal Employer-Household Data}
\acrodef{NACJD}{National Archive of Criminal Justice Data}
\acrodef{NDA}{non-disclosure agreement}
\acrodef{NLSY}{National Longitudinal Survey of Youth}
\acrodef{PAP}{pre-analysis plans}
\acrodef{PII}{personally identifiable information}
\acrodef{PSID}{Panel Study of Income Dynamics}
\acrodef{RADC}{restricted access data center}
\acrodef{RCT}{randomized control trial}
\acrodef{ReStud}{Review of Economic Studies}
\acrodef{TIER}{Project TIER (Teaching Integrity in Empirical Research)}
\acrodef{WRDS}{Wharton Research Data Service}
\acrodef{ZTRAX}{Zillow Transaction and Assessor Dataset}

\journal{Journal of Econometrics}

\begin{document}

\begin{frontmatter}



\title{Reproducibility and Transparency versus Privacy and Confidentiality: Reflections from a Data Editor}


\author[inst1]{Lars Vilhuber}

\affiliation[inst1]{organization={Cornell University},
            addressline={352 East Ives Hall}, 
            city={Ithaca},
            postcode={14853}, 
            state={NY},
            country={USA}}

\begin{abstract}
Transparency and reproducibility are often seen in opposition to privacy and confidentiality. Data that need to be kept confidential are seen as an impediment to reproducibility, and privacy would seem to inhibit transparency. I bring a more nuanced view to the discussion, and show, using examples from over 1,000 reproducibility assessments, that confidential data can very well be used in reproducible and transparent research. The key insight is that access to most confidential data, while tedious, is open to hundreds if not thousands of researchers. In cases where few researchers can consider accessing such data in the future, reproducibility services, such as those provided by some journals, can provide some evidence for effective reproducibility even when the same data may not be available for future research. 
\end{abstract}




\end{frontmatter}


\section{Introduction}
\label{sec:introduction}


In what follows, I will reflect on various aspect of the tension between reproducibility and transparency on one hand, and privacy and confidentiality on the other hand, from the perspective of a data editor who has had the privilege of observing the challenges, and some of the solutions, first hand. 
I will start by defining replication and reproducibility more precisely, since the terms are not always used in the same and unique way, and understanding which is which is important for what follows. Data access, or lack thereof, is key to the entire discussion, and I will briefly clarify that as well.  I will provide examples of situations that may arise for a variety of data access conditions, some of which may be surprising to readers. The examples are not chosen randomly, but do reflect, in my opinion, typical situations corresponding to frequently articulated concerns. I will then provide some speculative comments on where the economics and econometrics communities might head.

My focus will be on three main concerns, which I have often heard expressed by researchers. First, and this is where I will spend most of my time, I often hear that private, proprietary, or confidential data are an impediment to reproducibility, with the corollary that only ``open data,'' typically understood to be freely downloadable data, can be a part of reproducible research.  ``Confidential data'' is often equated to proprietary data. I will describe why that is incorrect, and why that confusion adversely affects the perception of reproducibility. Second, I will address the notion that access to data are the only barriers to reproducibility. I will highlight that access to other inputs to the production of research outputs may be equally constraining. Finally, I will briefly investigate the idea that data are infinitely re-usable, an argument that again needs to be unpacked in order to better understand the legitimate concerns expressed by both users of data, and guardians of sensitive data. 

I will provide arguments that contextualize these concerns. First, I will argue  that proprietary and confidential data can be, but need not be, part of fully reproducible research. This reproducibility does not come about automatically any more than it does in research that uses openly accessible data. Certain conditions must hold. My key concern, based on the work that we've been conducting at the \ac{AEA}, is not the fact that data may be confidential, but rather that some confidential data may only be accessible to a single researcher (or a single research group). Thus, it is the scale of access that is the characteristic of interest here, for which ``public-use'' and ``confidential'' are only very rough approximations. I will argue that reproducibility of the access protocols is crucial, and that transparency is key to addressing that concern. Second, when framing accessibility in this way, access can be viewed as a scarce resource. However, it is not the only scarce resource. Other barriers to reproducibility, and the general production of scientific knowledge are emerging --- or may have always existed -- with access to technical knowledge and sufficient computing resources being the important traditional dimensions. And thirdly, I will argue that while  data is technically reusable, each (new) use of data consumes some amount of privacy of the contributors of information captured in the data. Thus, privacy is also one of those scarce resources, and while it remains an open research and policy question how to measure and allocate the consumption of privacy, we should recognize it as something that will affect reproducibility. I will argue that more careful treatment of the data, in ways that we mostly know how to handle, can alleviate some of these concerns. 


\section{Some Definitions}
\label{sec:intro2:definitions}

Let me clarify how I use the terms ``reproducible'' and ``replicable'' in this article. One set of generally accepted definitions comes from  \citet{Bollen2015url} and \citet{national_academies_of_sciences_engineering_and_medicine_reproducibility_2019}: (computational) reproducibility is ``obtaining consistent results using the same input data, computational steps, methods, and code, and conditions of analysis'' \citep[pg. 36]{national_academies_of_sciences_engineering_and_medicine_reproducibility_2019}. Replicability is defined as  ``obtaining consistent results across studies aimed at answering the same scientific question [...] [obtaining] consistent results given the level of uncertainty inherent in the system under study'' (\textit{ibidem}).\footnote{The terms are now mostly used in this sense in the  social science and statistics community, but usage elsewhere may be the exact opposite \citep[e.g.][]{heroux_editorial_2015}, so readers should take care to identify the particular usage they face when reading the literature.} These definitions matter, because the narrower ``computational reproducibility'' is of lesser interest in furthering knowledge than ``replicability,'' yet it is the former that can be tested within a single manuscript, whereas the latter mostly arises from a literature. In the context of confidential data, this is particularly important, since reproducibility may be hard to verify on a particular confidential dataset, yet replicability of the basic scientific findings can be verified across many distinct and mutually exclusive confidential datasets.

Defining ``open data'' is in principle easy, but often misunderstood. Just because one can see it on the internet does not make data open. For instance, the S\&P~500, seen all over the internet, is not ``open data'' because it does not allow for free re-use. The Open Knowledge Foundation defines ``open'' as requiring machine readability, free of licensing charges, and free re-use, but does not mandate availability via download on the internet, absence of all fees, nor absence of any technical measures, such as a requirement to register and agree to abide by these rules \citep{openknowledgefoundation2015}. The United States has traditionally put its data products in the public domain (i.e., without any restrictions on usage or attribution). Many countries have switched their government data to default to open data \citep{statistics_canada_statistics_2012,uk_government_open_2014}. Well known ``public-use'' data are not always ``open'': IPUMS has a redistribution restriction (encapsulated in "Terms of Use", not a license), and some geographic data by international statistical offices remain under more stringent licensing requirements in other countries (e.g., United Kingdom). Many well-known surveys impose redistribution and usage restrictions, as I will discuss in one of the examples later. The interaction between licensing, openness, and re-use has ramifications that go beyond the scope of the present article. Good guidance on licensing of scientific data by its creators can be found in \citet{stodden_enabling_2009}.



\section{Confidential data and transparent research}
\label{sec:part1}

How can proprietary and confidential data  be part of transparent and accessible research, and how do reproducibility and replicability relate to that? Generically, we want to have confidence that the results in a manuscript were actually produced by research described in the manuscript. In the words of the slogan popularized by \citet{buckheit_donoho_1995}, ``a scientific publication is [...] merely advertising of the scholarship: [...] the complete software development environment and the complete set of instructions which generated the figures.'' Usually, however, the assurance that the results in the manuscript were obtained from the (provided) code and the underlying data is achieved by quite pedestrian means: By readers or, in some cases, data editors or computational reviewers, running the analysis themselves, i.e., by verifying computational reproducibility.%
\footnote{As a side note, a less frequently cited part of \citet{buckheit_donoho_1995} is their praise for the (at the time) novel availability of commercially provided ``quantitative programming environments'' --- \textsc{Matlab}, \textsc{Mathematica}, and \textsc{S-Plus}. Interestingly enough, much of the discussion 20 years later is around open source and free programming environments such as Octave, Python, R, and Julia as further enablers of computational reproducibility.}
The increasing use of active assessment of computational reproducibility at journals \citep{ChristianInt.J.Digit.Curation2018, VilhuberAEAPap.Proc.2020} is one way to render such checks more efficient, by signalling in the manuscript or the journal web pages that such an assessment has been conducted and found to be satisfactory. However, computational reproducibility is a very narrow criterion, and I postulate that just as important as the actual verification of the tables and figures are two additional corollaries. First, by successfully reproducing the tables and figures in the manuscript with provided code and data, the \textit{completeness} of the provided materials is verified. If either code or data were incomplete, some tables or figures would not be reproducible. Second, and I believe the most important corollary, once it has been ascertained that all materials have been provided, and the provided materials do, in fact, reproduce the results in the manuscript, it becomes much more reasonable to attempt to replicate the results by varying data inputs, different methods, alternate counterfactuals, etc. If the results were not reproducible, why would one  bother to attempt to replicate it? 

So what if one of the ingredients --- typically, the data --- is inaccessible to the reader? This is usually the complaint for ``proprietary'' data, understood to be microdata from companies or administrations. But the absence of data is not unique to such confidential data, nor is the absence of data in journal-hosted replication packages necessarily an impediment for conducting such exercises. I will proceed by using a set of examples (see Table~\ref{tab:sources}), most of which will be recognized by most economists as being of the ``confidential'' type. 

\begin{table}[htb]
\centering
\caption{Data sources discussed in main text\label{tab:sources}}

\begin{tabular}{rl}
\\
\toprule
1 & Compustat business information\\
2 & ZTRAX real estate transactions\\
3 & Ebay transaction data\\
\midrule
4 & Energy Information Agency data\\
5 & Wage records for WA state\\
6 & Swedish prison records\\
\midrule
7 & French customs records\\
8 & German Socio-Economic Panel\\
9 & National Longitudinal Survey of Youth\\
\bottomrule
\end{tabular}
\end{table}

Compustat is a widely-used dataset on business characteristics, currently produced by Standard and Poor's. It is accessible to many North American economists via the \ac{WRDS} service at the University of Pennsylvania.\footnote{\url{https://wrds-www.wharton.upenn.edu/}} It is proprietary data, and a subscription to \ac{WRDS} and a data request are needed. 
The \ac{ZTRAX} is composed of real estate-related records. Access was granted through a contract between the researcher's (academic) institution and Zillow, for free. Applications were accepted on a quarterly basis. 
Ebay is an ecommerce company, and its transaction data are a highly confidential data asset for the company.
Wage records of the kind that Washington State offers have been used for a long time \citep[e.g.][]{abowd_individual_1999}, through a wide variety of access mechanisms, including via the \ac{LEHD} data in the \ac{FSRDC} \citep{abowd_lehd_2009}. 
Wage records are just one kind of administrative records. Others might include energy production records from the U.S. \ac{EIA}, Swedish prison records or French customs records. 

Usually, survey data are not put into the same basket as ``proprietary'' data, and the presence of the \ac{GSOEP} and the (U.S.) \ac{NLSY} in Table~\ref{tab:sources} may surprise some readers. I will discuss both of these datasets here because they illustrate the boundaries of access that apply even to (relatively) open datasets.  

The datasets listed in Table~\ref{tab:sources} are not chosen randomly. Rather, they are mentioned here because they are a mix of frequently used data (Compustat, ZTRAX, wage records, NLSY, GSOEP) and others that are (so far) quite unique in their use in economics publications. They all have one thing in common: All were data inputs to manuscripts conditionally accepted in 2019-2021, during my tenure as the \ac{AEA}'s inaugural data editor. My team and I therefore scrutinized the replication packages in each of these cases for conformity with the \ac{AEA}'s data and code availability policy \citep{10.1257/pandp.110.dcap}. The policy requires that ``data and code used in the analysis are clearly and precisely documented and access to the data and code is non-exclusive to the authors.'' In a subtle change from prior years, authors are not exempted from the policy if their data is confidential; they are simply exempted from the requirement to provide the data to the \ac{AEA} as part of the publication of the data package. However, in almost all cases where data cannot be published, the data editor will make an attempt to either access the data, or  find a trusted third party who can do so in a timely fashion. In a nutshell, we verify that all of the data provenance is as clearly documented as can be. And then we test whether we can access the data, i.e.,  whether the data access procedure is reproducible. When that fails because of a lack of time, funding, or contractual complexity, we may turn to a potential trusted third party that does have access. However, if it fails because the procedure is not well documented, authors have to revise their instructions. Once data have been accessed, a reproducibility check is conducted,  and any computational issues  that may arise are noted.

In the year preceding this article, from December 2020 through December 2021, we ran reproducibility checks for 415 manuscripts. Of those, 11 manuscripts were outsourced to third parties, not always for reasons related to confidential data \citep{vilhuber_report_2022}. Since we started the active verification of replication packages, we have  cumulatively conducted over 1500 reproducibility assessments for over 1000 manuscripts, including the manuscripts that used the datasets mentioned in Table~\ref{tab:sources}. 

\begin{figure}
    \centering
    \includegraphics[width=0.8\textwidth]{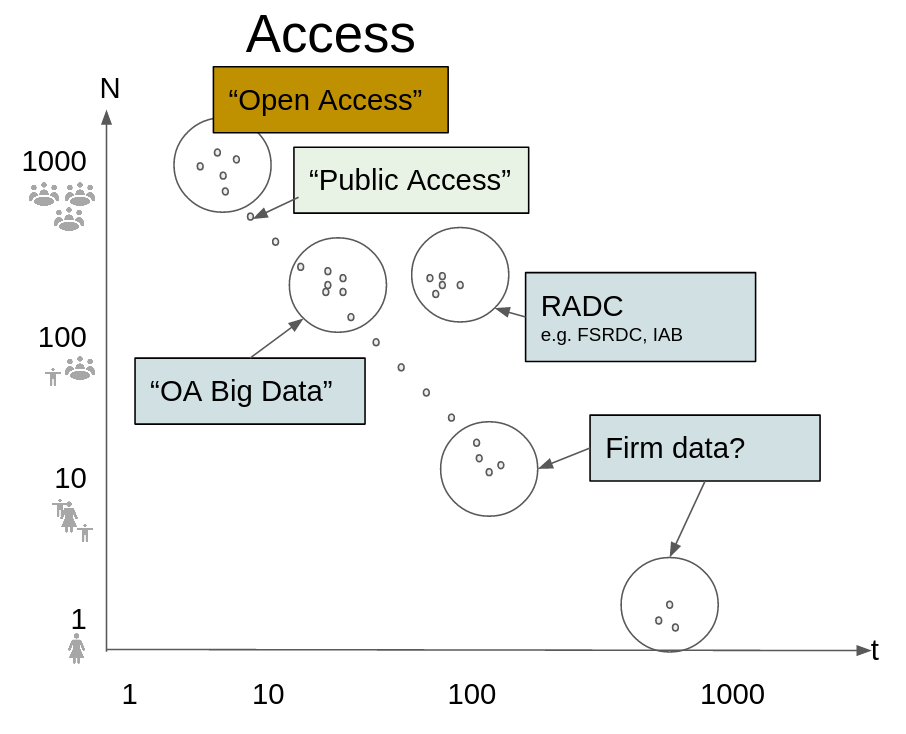}
    \caption{Conceptual trade-off between number of individuals accessing data, and time required to do so}
    \label{fig:nxt}
\end{figure}

So how do we assess that ``access to the data [...] is non-exclusive to the authors?''  Access is not necessarily a single-attribute concept. Open data generally means that access is by anybody (who, how many), from anywhere (without restrictions), by simple download (how easy), for free (what cost). Restrictions, cost, and access mechanisms are obstacles that can always be overcome, given time: grants can be written, contracts signed, proper programming or storage obtained. Thus, time is a sufficient statistic for many factors, which I conceptually correlate with the number of individuals able to access it (see Figure~\ref{fig:nxt}). Open access data may be accessible to thousands of individuals, though when the data become ``big'', that accessibility may decline --- think of downloading a terabyte of satellite image data, even if it is open data. Data that the public can access, but which is not simply downloadable --- for instance data obtained through freedom of information acts or ``sunshine laws'' --- will also reduce the number of people willing to go through the process. Confidential data accessible through \acp{RADC} such as the U.S. \ac{FSRDC}, but also the French \ac{CASD} or the research data center of the \ac{IAB}, is usually accessed by hundreds of researchers and potentially accessible to even more, despite lengthy and often complicated access procedures. As of October 2021, there were 1,042 researchers working on active \ac{FSRDC} projects using Census Bureau data \citep{us_census_bureau_center_2021}, and 937 approved projects in the \ac{CASD} system \citep{centre_dacces_securise_aux_donnees_projets_2022}. Firm data such as Compustat may have several hundred researchers over the years, whereas more recent data such as ZTRAX might be in the dozens of researchers. Of course, there are numerous datasets, such as the Ebay transaction data, that have only been accessed by a handful of researchers.

Positions in this conceptual graph are not fixed over time. Data have migrated into RADCs, becoming more accessible to more researchers over time (shifting towards the top left), but data have also been de-accessioned, or access procedures shut down, rendering access impossible or at the very least, extremely lengthy (shifting horizontally to the far right). 

Clearly, the typical proprietary dataset provided by a private company is accessed by far fewer people than the typical public-use survey dataset. It takes a long time to put together proposals, relationships, whatever is needed to do that. So can access to firm data be made easier, to the point of being reproducible? This is where the quality of the data access documentation from researchers in their replication packages is critical. The second time that somebody wants to access the data from the same firm should not require reinventing the wheel. Thus, assessing the feasibility of such a data access reproduction is part of the data editor's tasks. Other efforts to streamline such access work through institutional access, such as \ac{WRDS} or the Private Capital Research Institute \citep{JengAm.Econ.Rev.2016,lerner_private_2021}, expanding access and making research relying on these data more reproducible.\footnote{For several other examples, see \citet{cole_handbook_2021}.}

Without specifying a formal model, one way to think about what is needed to induce researchers to generate  reproducible, credibly transparent manuscripts and replication packages is a threat of entry model. If somebody else can access the same data the original researcher did, armed with the same code, the same results should be obtained. In the fully open access model for both data and code, this is likely the case. When only the original researcher can ever access the original data, the threat of entry does not bite. Where is the boundary, how high is the minimal level of feasible entry --- access to the data --- that induces the correct behavior? One hypothesis is that at least one other, independent and unbiased researcher is enough. Data editors conducting pre-publication reproducibility checks is one option, but it need not be the only option. But in order to ascertain --- publicly and openly --- what the number could be, data access documentation is key. Access itself must be reproducible, repeatable, by individuals other than the original author. But what are the conditions for such access to be feasible? 

Of course, the ultimate goal remains replicability -- application in a different context -- and generalisability.  If I had transaction data from eBay, what do I need to do to obtain transaction data from some other online marketplace? If I had access to wage records  in Sweden, can I reproduce access to (similar) wage records in Italy, Taiwan, or in the United States? In order to do so, the description of access may be less useful, but description of salient data characteristics, of data processing steps, becomes relevant, in particular when the original confidential data are not broadly accessible. 

%

In my work for the AEA, it is standard practice to at least attempt to access the confidential data that are used by manuscript, though this does not always succeed in a timely fashion. However, we are successful sufficiently often, even in tricky cases. For instance, in 2021, we signed \acp{NDA} or \acp{DUA} for 48 of the manuscripts that we assessed. I mention all of the datasets listed in Table~\ref{tab:sources} because  we have, in fact, encountered them all as part of our work. 

\paragraph{Compustat} We have accessed and retrieved Compustat data multiple times, thanks to my university's subscription to \ac{WRDS}. We also have an agreement with Standard \& Poor's, creators of Compustat,  that allows authors to directly share  data extracts with us, because we are a subscriber.  Compustat is not open access data, yet it can still be accessed by thousands of researchers world wide who are paying subscribers. 

\paragraph{ZTRAX} Zillow provided \ac{ZTRAX} data free for academic research, using a relatively simple access procedure. An \ac{NDA} was signed by the university, and applications were reviewed by Zillow once every quarter. Thus, while there may have been a short delay, potential access was very broad. Yet, ZTRAX data are  not open access, and remain off-limits to non-academic researchers. Furthermore, and with ramifications which we discuss below, access to this proprietary dataset has been shut down. As of July 1, 2021, no further access requests are accepted, as Zillow has shut down the ZTRAX program, but existing contract holders can continue to work on \textit{existing} approved projects \citep{zillow_zillows_2021}. 

\paragraph{eBay} We signed an \ac{NDA} for eBay transaction data for one paper \citep{10.1257/aer.20190460}. The contract and contact were facilitated by the authors, but our access was unrestricted and independent. Whether others will  be able to access the data in the future is, of course, a different question. 

\paragraph{EIA} EIA data access procedures were paused around 2016, and, as of the time of writing of this article, have not resumed. Nevertheless, prior to that date, many researchers in energy policy and industrial organization were able to access the data. In fact, because they continue to have access to the data, they sometimes have well-documented disagreements about inference and analysis of the data \citep[see][]{cicala_when_2021,han_when_2021}. 

\paragraph{Wage records} Administrative data on pay were some of the earliest to be used, whether at the national \citep{irelan_retirement_1988} or the state level in the US \citep[e.g.][]{abowd_individual_1999}, and elsewhere (Scandinavia, Germany, France, Brazil, Canada, to mention  those used in a forthcoming special issue of this journal). They are obviously confidential (except in a few exceptions, such as  federal employees in the United States), but they have been accessed by thousands of researchers, working through restricted access environments such as the \ac{FSRDC} in the US, \ac{CASD} in France, as well as individual \citep{hawley_ohio_2021} or institutional arrangements. We regularly verify the reproducibility of Brazilian, German, and some Scandinavian data, though our access to Canadian and (ironically) US data is somewhat more restrictive. 

\paragraph{Prison records} For \citet{10.1257/app.20200615}, the authors arranged for me to have access to Swedish prison records, working on a remote Swedish server. However, in this case, another one of the regularly occurring barriers came into play: in order to access the data, I needed to be physically in Europe. My access was coordinated with personal travel to Europe, however, when that travel was cancelled due to the Omicron wave of COVID-19, I was not able  to follow through. We did, however, prove that access was feasible. Closer to home, many researchers were stymied in their access to computing resources when COVID shut down in-person access to the \ac{FSRDC}. Nevertheless, in normal times, thousands of researchers could, in principle, access the Swedish prison data, and can, in principle, access data in the \ac{FSRDC}. In a case using prison data from Mississippi, we worked with a graduate student who had access to the data via the author. While not quite as arms-length as a fully independent access mode, we nevertheless were able to verify reproducibility, and correct some minor issues \citep{mukherjee2021}.

\paragraph{French customs records} Often, the two access modes --- individual arrangements with data providers, and structured access through restricted access data centers --- overlap. For \citet{piveteau2021}, the author had originally accessed the data through a private arrangement with the French customs agency. However, by the time the manuscript was submitted to my team, the data were available through the \ac{CASD} --- not just to my team, but to any future replicator. We worked with a replication service called \texttt{cascad} \citep{perignon_certify_2019}, who were able to access the data and run the author's code. Although the data were notionally the same, the two extracts --- the author's and the one accessible via \ac{CASD} --- were subtly different. The reproducibility verification then turned into a replication and robustness exercise, and ultimately was documented in the manuscript's appendix.\footnote{See \url{https://perma.cc/BYF4-NN2H} for more details.} The broader message here is again one of broadening of access: hundreds of researchers in Europe have access to \ac{CASD}, and at least in some cases, research can even be conducted by researchers located in North America, remotely and securely accessing confidential French administrative data. Similar access is feasible for German data through the \ac{IAB}.

\paragraph{Survey data} Most researchers assume that the most ``publicly accessible'' datasets such as the \ac{PSID} or the \ac{GSOEP} are easy to access. But the changing landscape of privacy and confidentiality make that assumption fail. The PSID is not, in fact, redistributable, despite multiple replication packages in economics containing extracts from the PSID (infringing on the terms of use the authors originally agreed to when downloading the data). In order to access the \ac{GSOEP}, users outside the European Union must have their university sign various legal agreements required by the data producer's interpretation of the \ac{GDPR} \citep{gdpr_2016}. These agreements are not just \textit{pro forma} agreements --- in fact, my university as well as several others have not been able to agree to sign these forms for more than a year (and as of February 2022, still had not signed).  \ac{GSOEP} are widely used, including by many researchers in the United States.\footnote{See for instance \citet{wagner_english_1993} and the subsequent evolution of the Cross-National Equivalent File \citep[\url{https://www.cnefdata.org/}]{burkhauser_cross-national_2000}.} Thus, the same dataset that I could access without issue while writing my thesis more than twenty years ago from Canada was now inaccessible to me from the United States. Note that even if I had accessed the data, it would not have been the same data as that accessed by European researchers. 

Conversely, most of the \ac{NLSY} data can be freely downloaded by any user worldwide --- an excellent example of open access data in the public domain --- but precise geo-located data is, for privacy reasons restricted. Less well known is that the restrictions are more than a simple (free) application: the restricted-access data for the \ac{NLSY} can be accessed by US researchers only, and remains inaccessible for non-US researchers \citep[see, for instance, the appendix to ]{mogstad2021}. We, on the other hand, were able to access the data ourselves, and verify computational reproducibility. In addition, we were able to request that a replicator within the Bureau of Labor Statistics run the code for us, verifying the results.

A common thread through these several examples is that more than one person, and sometimes several thousand researchers, were in principle able to access the data. While we took the first step for a particular paper, we are unlikely to be the last to take a paper with restricted-access data, and assess reproducibility, replicability, and generalisability. In all cases, even when we failed to access the data, as in the case of the \ac{GSOEP}, we were able to improve the documentation of how others may access the data, having tested the mechanisms ourselves. The access to any of these data, whether now or in the future, has thus been made more transparent.

In sum, I have argued that open access data is not a necessary condition for reproducible research, illustrated that  proprietary and confidential data can be used for fully reproducible research, and have pointed out the importance of transparency of the access procedures and overall accessibility of data for broader replicability and generalizability of applied research. Risks to replicability remain:  Degradation of access to data over time is the most important, though it is not unique to confidential or proprietary data. Reproducibility assessments close in time to when the original research was conducted may mitigate some concerns about the reliability of research with any data. 

Clearly, fully open access data, properly preserved and curated, leads to the greatest ease, dispersion, and persistence of reproducibility. But evolving rules and standards surrounding privacy and confidentiality may lead to legitimate restrictions on various data. Reproducibility and replicability are not a dichotomous characteristic. As data access stops being fully open, assessing reproducibility and implementing replications becomes ever more challenging, but rarely impossible.

\section{Confidential data, proprietary data, and access data}
\label{sec:conf-prop-data}

This brings me to my second theme: Confidentiality is only weakly correlated with the proprietary nature of some data. All data are created by a data creator with a specific purpose in mind. The data creator is thus the first data custodian: Tasked with guarding, protecting, and maintaining the data. Researchers  or research groups conducting surveys are tasked as data custodians by the ethic boards or \acp{IRB}  that approved the activity. Government agencies that collect administrative data (EIA) or conduct surveys (BLS, Census Bureau) are mandated by law to protect the privacy of respondents. Private companies collecting data from and for clients may have some legal obligations to protect the clients' or consumers' privacy, with varying compliance. Each of these various types of data custodians has a different relationship between the data collected and the primary business model. For survey data collected by researchers and government agencies, the typical primary purpose is to make the data available to a broader audience. Government agencies often make the data available in raw form, whereas researchers may prefer to make the information gleaned from surveys available processed as scientific articles. Data collected by private companies and administrative data collected by government agencies is not primarily collected for the purpose of publishing it, and thus is often subject to other constraints. Researchers and private companies in particular are quite similar in focusing on their primary purpose, transforming the data into their principal output --- a commercial product or a scientific article. The main difference is that research funders and journals, through data availability policies, may be able to coax researchers into making the underlying raw data available, whereas such inducement mechanisms have little hold on private companies. 

Thus, balancing the competing objectives of primary data collection purpose, privacy concerns, and data availability, is the challenge that data custodians face. Even when data collection purpose and data availability are aligned, privacy concerns play a role, ultimately defining how many researchers (or citizens) may access the data. In all cases, the data custodian makes exclusive decisions about access - and thus is a ``proprietor'' of the data in one of the senses of the word.\footnote{Merriam Webster defines ``proprietary'' as ``used, made, or marketed by one having the exclusive legal right.''} A researcher conducting a survey is a ``proprietor'' of the data collected, just as Zillow is the (private) ``proprietor'' of the property transaction data it has collected. The Census Bureau is the ``proprietor'' of the data collected via its decennial census. The proprietors differ not so much in how they can decide to give access to the data collected, but differ much more in the primary objective of their data collection activities, and the constraints and obligations this may impose on making the data available.

Of course, the private owners of  data --- by another definition of the word, ``proprietary'' data\footnote{The American Heritage dictionary defines ``proprietary'' as ``privately owned, as a business.''} --- are likely to be more restrictive in providing access. But much of the historical discussion about data availability has also focused on inducing (or forcing) academic researchers to relinquish exclusive (proprietary) access to their research data. 

I would thus argue that a more useful discussion is about \textbf{access}. The most comprehensive, unbiased, and robust access models are used by government agencies. As public-use data under an open data license \citep{uk_government_open_2014,statistics_canada_statistics_2012} or by putting data into the public domain, as in the United States, these models have defined the most open process, to the benefit of both researchers and society more broadly \citep[f.i. see][on benefits of public domain satellite imagery]{nagaraj_improving_2020}. But they also maintain robust restricted data access centers which, despite their many shortcomings, make data much more accessible (\ac{FSRDC}, \ac{CASD}, and many more). Such data access centers can also work for privately-owned data \citep{weinberg_science_2014,lerner_private_2021}. Other options include data intermediaries, or data trusts \citep{open_data_institute_what_2018}. Encouraging is that many private providers increasingly report dedicated units for academic cooperation \citep{harris_understanding_2017}. Greater information and transparency will reduce the burden of requesting access, in particular when combining with demonstrating trustworthiness among a broader set of researchers, such as legally specified ``accredited researchers'' \citep{uk_government_digital_2020} or via a ``researcher passport'' \citep{levenstein_researcher_2018}. 

In the meantime, transparency of the access procedures through author-provided documentation is the most important element. This includes clearly describing restrictions on data access: whether only three people can access it, whether there is a citizenship or residency requirement, whether access will terminate one year after publication, or in five years. The template README we have developed \citep{vilhuber_lars_template_2020} helps authors describe such access procedures and restrictions. 

I do note that amongst the examples I described in the previous section are cases that illustrate the unpredictability of access when data and their access protocols are not properly preserved, or when unforeseen legal barriers arise. As researchers, we often take for granted that openly accessible data are easily accessed not just now, but also in the future. Data subject to an open license, but posted on a website or Github repository, are not much better in this regard. Websites or Github repositories can be deleted at any time by the owner, or disappear for other reasons. While copies may exist, for instance tucked away in replication packages, the data are no longer easily accessible. Only when preserved at trusted repositories can persistence over long periods of time --- euphemistically referred to as ``forever'' --- be assured. In some cases, data are known to be destroyed after a fixed period of time, e.g., ``one year after publication.'' But consider the ZTRAX, EIA, and GSOEP cases.  In the case of ZTRAX, all historical data as well as current data are disappearing, because the data provider is no longer willing to provide a (free and voluntary) service. In the case of the EIA, current data are also no longer made available to anybody (for now), but historical data can be accessed by various research groups that hold valid (indefinite length) contracts. Could such data, soon obsolete for commercial purposes, be made available in perpetuity, even if under certain conditions, and at no (or lower) cost to its original data creator? Emerging structures such as ``data trusts''\footnote{Data trusts ``involve one party authorising another to make decisions about data on their behalf'' \citep{open_data_institute_defining_2018,open_data_institute_what_2018}, though most discussion seems to be about contemporaneous access, not future access to archival data.} may help, but are rare. For now, the decision by Zillow to sunset the ZTRAX data program is surely unforeseen by many researchers. Replicators will no longer be able to access the data in the future, even though the data were broadly accessible at the time that articles used the data.

Our difficulties in accessing \ac{GSOEP} data, and every researchers inability to access EIA data, are  examples where access may be curtailed through events outside of the control of the data producing entity. While many thousands of researchers within the European Union continue to be able to access the GSOEP data, access outside of the EU is curtailed. Access to EIA data may yet return, subject to various political and budgetary decision processes. However in contrast to the ZTRAX situation, these impediments may only be temporary. But they are a warning that even broadly  accessible data, even when hosted and properly curated by an established data producer (in the case of the GSOEP), may become inaccessible in the future.

\section{Other factors constraining reproducibility}
\label{sec:other_factors}

Implicit in Figure~\ref{fig:nxt} is the idea that a \textit{threat} of entry keeps researchers honest. If others can later come in and verify a researcher's work, the work will be made to be reproducible. How many potential entrants there need to be remains undefined. By providing at least one high-likelihood entrant prior to acceptance, journals' data editors can play a part, but post-publication verification and replicability also have a role to play. Reducing barriers to data access, by increasing the number o potential entrants, thus has the potential to increase intrinsic reproducibility.  However, barriers to entry  are not specific to data. Other inputs to the research production process are software, computer code, and computing resources. Software in this context is typically a ``quantitative programming environments'' \citep{buckheit_donoho_1995} (e.g., MATLAB, Stata, or the Intel Fortran Compiler), and computer code are the high-level instructions within such software (MATLAB, Stata, or Fortran (source) code).  While data (and code) availability policies can largely handle the problem of availability of computer code, both software and computing resources can be additional constraints. 

Software resources can be constraining for some researchers. Much of economics research relies on commercial closed-source software: Stata was used in 73\% of the 2,552 replication packages archived by the \ac{AEA} between 1999 and 2019 \citep{10.1257/pandp.109.718}, followed by MATLAB (22\%) and SAS (4\%). Licenses for these software may not be available to young researchers, researchers in emerging economies, or to non-academics. Within a certain field, this is likely to be less of a problem, yet it should be kept in mind. Even when conducting reproducibility assessments at one of the top universities in the United States (Cornell University), my team regularly encounters software licensing issues that prevent us from conducting a full computational reproducibility check. 

A more complex issue are compute cycles and storage, i.e., computing resources. In 2021, 3\% of replication packages assessed by my team were larger than 20 gigabytes \citep{vilhuber_report_2022}, excluding several packages in the 100 gigabyte range, which were archived elsewhere. This also excludes packages that require the download of data from non-redistributable sources or which are already archived elsewhere, which are known to add up to 30 additional gigabytes. Such data must be downloaded, locally stored and processed, within a reasonable timeframe.  Having enough time to acquire all the necessary data is a constraint on our replication process, on the reproducibility checks, but also for future access by others. 

But the same is true for computing resources. Whether a replication package requires \ac{HPC}, high throughput computing, or scalable computing, any replicator must have the ability to access such resources within a reasonable time, plus the time to then let the code run. Running code within a limited set of quantitative programming environments is de facto standardization that greatly facilitates reproducibility. Such standardization is mostly absent from the way most computational or social scientists leverage computing resources beyond their workstation or laptop. There is no unified approach to computing in either university \ac{HPC} environments or commodity commercial cloud environments. No institution, even large universities, do not have billing or access arrangements set up across all commercial cloud environments. 

Researchers bear some of the burden. Most researchers rely on local IT staff for their own access, yet are often challenged to describe exactly how that access works so that others, at other institutions, can reproduce it. 
Even when all the pieces come together, access to \ac{HPC} resources is rarely free. When researchers state that code requires ``20,000 core-hours'' \citep{rudik_optimal_2020}, both duration and the cost of the computations constitute a barrier for both data editors and future replicators. 

Authors of replication package instructions also rarely take into account the distinction between computational reproducibility --- the code and parameters that exactly reproduce the results in the paper ---  and broader replicability. For instance, guiding a replicator to reduce the number of  bootstrap replications from, say, 10,000 to 100 may allow that replicator to vary other elements of a computational model, and better assess the robustness of the model during a replication. A good replication package should assist a future researcher in their replication attempts, not just their computational checks. 

As mentioned before, many of these constraints can be lifted given enough time. Grants can be written to fund data access and computing resources, IT help enlisted, data downloaded, specialized software purchased or programmers hired to rewrite in open source software code. Time is sufficient statistic for most of these constraints.

\section{Why data are  not infinitely re-usable, or why we consume too much privacy}
\label{sec:resuse}

Finally, one additional constraint ties together the two previous sections. Privacy can also be a limiting factor, and is not well captured by time as a sufficient statistic. If we consider confidential, and thus privacy-protected, data as a one-time resource, not accessible to others, then whatever analysis is conducted with the data may be allowed to ``use up'' as much privacy as it wants.%
\footnote{In the discussion here, I will consider the data as collected, and ignore the privacy trade-offs that may arise in data collection as well, see \citep{GhoshArXiv10111375Cs2011,LiProc.16thInt.Conf.DatabaseTheory-ICDT132013,couper2008,CouperJ.Off.Stat.2010} for some discussion.}
But when confidential, privacy-protected data are to be re-used,  privacy can become a constraint. This is currently mostly a theoretical concern, but more sophisticated privacy-preserving analyses \citep{DworkJPC2018,YuJ.Biomed.Inform.2014,LinCoRR2013} are making their way into economics, as illustrated by Friedman and co-authors \citep{ChettyJ.Priv.Confidentiality2019} and discussed in this same session.\marginpar{\tiny Should I reference Friedman in this session?}  But I will also point out that we may be needlessly ``wasting'' the scarce resource that is privacy.

\begin{figure}
    \centering
    \includegraphics[width=0.6\textwidth]{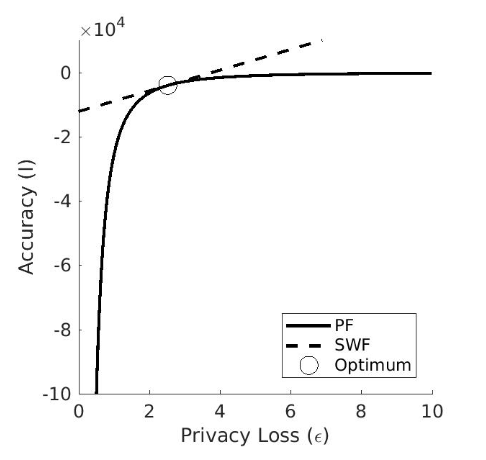}
    \caption{Privacy --- Accuracy Tradeoff}
    \label{fig:abowdschmutte}
    \footnotesize Reprinted from \citet{AbowdAm.Econ.Rev.2019b} with permission.
\end{figure}

\citet{AbowdBrook.Pap.Econ.Act.2015a,AbowdAm.Econ.Rev.2019b} discuss the trade-off between privacy and accuracy. For any given level of accuracy, there is a corresponding level of privacy that can be (probabilistically) achieved when using the best methods available (see their Figure~1, reprinted as Figure~\ref{fig:abowdschmutte}). While much of the discussion in \citet{AbowdAm.Econ.Rev.2019b} centers around the one-time release of data, the same basic principle applies to any analytical use of the data, which usually requires release of far smaller quantities of data. As \citet{AbowdBrook.Pap.Econ.Act.2015a} discuss and \citet{ChettyJ.Priv.Confidentiality2019} practice, analytical methods can be appropriately adapted to both the privacy-preserving methods used and to the chosen level of privacy protection applied to a particular application of the method. \citet{DworkJPC2018} discuss the problem of how to combine access to raw data with release of privacy-protected statistics, and discuss the idea of having $k$ partitions of the data sufficient to reliably choose a statistic when the statistic is unknown. Multiple authors have looked into sampling as an amplification mechanism for modern privacy-protection methods \citep{NissimProc.Thirty-NinthAnnu.ACMSymp.TheoryComput.-STOC072007,LinProc.2013Int.Conf.Manag.Data-SIGMOD132013,LiProc.7thACMSymp.Inf.Comput.Commun.Secur.2012} --- a method that has a long history in traditional disclosure avoidance measures as well. 

But many modern econometric analyses of population-size datasets use \textit{all} the data, and marvel at how small their standard errors often are. This suggests that the same analyses could have been conducted with a much smaller sample, with insignificantly weaker inferences, while using substantially less of the privacy budget \citep{Abowd13thBienn.Fed.Comm.Stat.Methodol.FCSMPolicyConf.2016}. How to do this prior to an analysis of the data is well known to those conducting surveys: conduct a power analysis. Compute the sample size necessary to make the desired inferences. Researchers would then obtain access to a subset of the data, chosen to be sufficiently large to permit the desired analysis, yet not spending too much privacy. Those researchers having to conduct their own surveys have long planned their surveys this way --- every additional respondent has a discrete monetary cost. Extending this thinking to the use of confidential data acknowledges that every additional observation carries a discrete \textit{privacy} cost. Clearly, such analysis-specific extracts will have additional implications for the reproducibility of research. Data custodians --- presumably \acp{RADC} in most cases --- will want to preserve such extracts, and make them available to those wishing to verify the exact computational reproducibility. Independent replicators will need to incorporate the uncertainty due to sampling and privacy-protection into their analysis.

\section{Outlook}
\label{sec:discussion}
\label{sec:outlook}

How do I expect that privacy, confidentiality, reproducibility, and more generally econometric analyses will interact in the near future? Mostly, of course, I speculate, or reflect on emerging projects or activities. 

First,  the growing tension between increased privacy concerns and feasible analyses will likely lead to changes in how data are collected and accessed. I would expect that, as outlined in the previous sections, data custodians may require more stringent justifications not just for access to the data, but also for the size of the requested sample, combined with newer methods of disclosure avoidance. As outlined in the previous section and by others in this session, such requirements need not impact researchers' ability to draw inferences from the data, but it will require a different type of planning, involving power calculation and something akin to a pre-analysis plan \citep{olken_promises_2015}. Why request a 100\% sample when a 1\% sample may suffice, and is much faster to compute with. Many \acp{RADC} already require detailed project proposals, sketches of proposed tables, and justification of why access to confidential data is necessary, so such new requirements would not be radically new. 

However, I also expect that the same privacy concerns will lead to less public-use data being made available using traditional disclosure avoidance methods, and some to switch to formal privacy methods instead. This may, but need not, impact the ability of certain analyses to draw inferences. At the same time, formally private methods allow for vastly richer tabular data to be released, as evidenced by some such recent releases \citep{foote_releasing_2019,ChettyJ.Priv.Confidentiality2019}, and by data releases that I have been involved with in the past \citep{abowd2009,MachanavajjhalaProc.2008IEEE24thInt.Conf.DataEng.2008}. Even when freely downloadable public-use microdata are made available, researchers will need to (more consistently) take into account the privacy protections applied to the data, as \citet{AbowdBrook.Pap.Econ.Act.2015a} have argued.  I note that many data resources created by private companies, such as Google's Community Mobility Reports \citep{aktay_google_2020}, \citet{safegraph_social_2020} Social Distancing Metrics, or Meta's Movement Range Maps \citep{dow_protecting_2020}, already use differential privacy, but I am do not know whether any of the hundreds of papers that use one or more of these datasets take the disclosure avoidance technique into account when analyzing the data and making inferences. Again, this is not radically new: researchers have long known to use weights reflecting sampling and design of surveys, and many canned econometric routines in popular statistical programming environments are available. I would expect that such privacy-aware canned econometric routines will also emerge,  allowing researchers to take into account various disclosure avoidance techniques associated with data releases. Data custodians are on the hook to provide more consistent metadata accompanying protected data releases to allow such routines to work seamlessly, even when the researcher is not an expert in differential privacy.

A potential reduction in new public-use microdata releases is generally not a good outlook, because the traditional model of accessing confidential data, via secure laptops, thin clients, or from within secure rooms \citep[see][for an overview]{shen_physically_2021}, is becoming increasingly obsolete. I do, however, expect a resurgence in ``remote job submission systems'' (though they may not be called that), where researchers prepare code outside of secure environments, and submit non-interactive code to a system that runs code and provides privacy-protected output back to the researcher. For instance, such systems are in currently in use for German (\ac{IAB}), Canadian,\footnote{\url{https://www.statcan.gc.ca/en/microdata/rtra}}, and Norwegian data.\footnote{\url{https://microdata.no/en/}} While such systems can never be a full replacement for interactive development of data analysis code, they are, or can be made to be, sufficiently powerful to replace many types of analyses. Newer types of systems might use synthetic or test data (as the \ac{IAB} already habitually provides), differential privacy \citep{barrientos_providing_2018}, or containers. I would expect that some such access mechanisms are integrated into software packages, in ways quite similar to how various software packages already exist to access data through \acp{API} \citep{fredr,tidycensus,statafred} and in cloud services \citep{analogsea,heiss_create_2018}. In addition, such methods might allow for more robust inferences, because the analysis is ``blind'' \citep[][see also \citealp{dwork_preserving_2016,dwork_differentially_2021}]{maccoun_blind_2015}.

I also think that respondent consent, in particular influenced by \ac{GDPR} and its California counterpart, can and should become more fine-grained. Rather than having a blanket consent (or not) on survey response, respondents may specify what type of analysis could be conducted with their response to a particular item. Econometrically, this might not be too hard to accommodate --- item-level imputations come to mind --- but the privacy protection systems in place need to be able to handle such detailed permission schemata. Explicitly given consent for the research use of administrative data may not be far behind.

Finally, much of the research being conducted using confidential data and submitted to the AEA journals, as described in Section~\ref{sec:conf-prop-data}, has been verified to be by and large very reproducible, once access to the data has been obtained. However, what if authors could provide evidence of reproducibility that does not require a second check? Computer systems in \acp{RADC} or corporate systems could be configured to issue ``trusted computing certificates,'' tying together computer code, output, and data into a single verified run. Authors provide only computer code and output together with the certificate, without the data, but in doing so, provide sufficient transparency without revealing confidential data. Such evidence is implicit in the remote processing systems used by the \ac{IAB}, and is akin to the ``reproducible run'' provided by services such as \href{https://codeocean.com}{codeocean.com}. They could be combined with the software packages alluded to earlier to provide streamlined, efficient, and reproducible computing on confidential, even proprietary, data.




\appendix


 \bibliographystyle{elsarticle-harv} 
 \bibliography{extra,references-zotero,PrivacyBibliography}





\end{document}